\documentclass[12pt,preprint]{aastex}
\shorttitle{Metallicity gradients in early-type galaxies.}
\shortauthors{OGANDO ET AL.}

\begin{document}

%----------------------------------------------------------------------

\title{Do observed metallicity gradients of early-type galaxies
support a hybrid formation scenario ?\altaffilmark{1}}
\author{Ricardo L.C. Ogando\altaffilmark{2}, Marcio A.G.
Maia\altaffilmark{3,4}, Cristina Chiappini\altaffilmark{5}, Paulo
S. Pellegrini\altaffilmark{3,4}, Ricardo P.
Schiavon\altaffilmark{6}, Luiz N. da Costa\altaffilmark{3,7} }

\altaffiltext{1}{Partly based on observations at European Southern
Observatory (ESO) at the 1.52m telescope under the ESO-ON
agreement.}       \altaffiltext{2}{Instituto de F\'\i sica,
Universidade Federal do Rio de Janeiro, (RJ), Brazil;
ogando@if.ufrj.br}        \altaffiltext{3}{Observat\'orio Nacional
(ON/MCT), Rua General Jos\'e Cristino,~77,  Rio de Janeiro
20921-400 - RJ, Brazil; maia@on.br, pssp@on.br}
\altaffiltext{4}{Observat\'orio do Valongo, Universidade Federal
do Rio de Janeiro, Ladeira Pedro Ant\^onio 43, Rio de Janeiro -
20080-090 - RJ, Brazil}       \altaffiltext{5}{Osservatorio
Astronomico di Trieste, via G.B. Tiepolo 11, 34131 Trieste, Italy;
chiappin@ts.astro.it}         \altaffiltext{6}{Department of
Astronomy, University of Virginia, P.O. Box 3818, Charlottesville,
VA 22903-0818; ripisc@virginia.edu}   \altaffiltext{7}{Present
address: European Southern Observatory, Karl-Schwarzschild-Strasse
2, D-85748 Garching, Germany; ldacosta@eso.org}

%----------------------------------------------------------------------

\begin{abstract}
We measure radial gradients of the $Mg_2$ index in 15 $E-E/S0$ and
14 $S0$ galaxies. Our homogeneous data set covers a large range of
internal stellar velocity dispersions ($2.0 < log\ \sigma < 2.5$)
and $Mg_2$ gradients ($\triangle Mg_2/ \triangle log\ r/r_e^*$\ up
to -0.14 mag dex$^{-1}$). We find for the first time, a noticeable
lower boundary in the relation between $Mg_2$ gradient and
$\sigma$ along {\it the full range of $\sigma$}, which may be
populated by galaxies predominantly formed by monolithic collapse.
At high $\sigma$, galaxies showing flatter gradients could
represent objects which suffered either important merging episodes
or later gas accretion. These processes contribute to the
flattening of the metallicity gradients and their increasing
importance could define the distribution of the objects above the
boundary expected by the ``classical'' monolithic process. This
result is in marked contrast with previous works which found a
correlation between $\triangle Mg_2/ \triangle log\ r/r_e^*$ and
$\sigma$ confined to the low mass galaxies, suggesting that only
galaxies below some limiting $\sigma$ were formed by collapse
whereas the massive ones by mergers. We show observational
evidence that a hybrid scenario could arise also among massive
galaxies. Finally, we estimated $\triangle[Z/H]$ from $Mg_2$ and
$H\beta$ measurements and single stellar population models. The
conclusions remain the same, indicating that the results cannot be
ascribed to age effects on $Mg_2$.

\end{abstract}


%---------------------------------------------------------------------
%\vskip 1cm {\hskip 5cm \bf Version 2.10  20-07-2005}
%---------------------------------------------------------------------

\section{Introduction}

Two classical galaxy formation scenarios have been proposed to
explain the build-up of elliptical galaxies: the Monolithic
Dissipative Collapse (MDC) and the Hierarchical Clustering (HC).
The existence of metallicity gradients in galaxies provide a key
clue to help deciding between these two scenarios.

In the MDC \cite[e.g.][]{Egg62, Lar74, Chi02} a galaxy forms by
means of a rapid gravitational collapse, with a considerable
dissipation of energy from a cloud of primordial gas. During this
process, star formation occurs in a short period of time. If the
galaxy gravitational potential is strong enough to retain the gas
ejected by stars and supernovae, it migrates to the galaxy's
central regions. New generations of stars will be more metal rich
than those in external parts and a steep radial negative
metallicity gradient (MG) is established as a mass-dependent
parameter. In this scenario ellipticals form at high redshift and
on shorter timescales than spirals, and are assembled out of gas
and not of pre-existing stars. Observational evidence for high
mass galaxies already in place at higher redshifts have increased
in the last years \citep[e.g.,][]{Cim04, Gla04}. The appealing
aspect of such models for the formation of ellipticals is that
they can explain most of the observational constraints relative to
stellar populations. In particular, those assuming the inverse
wind hypothesis \citep{Mat94} can also explain the observed
increase of $\alpha/Fe$ with galaxy mass \citep[e.g.,][]
{Mat03,Tho02}(see also Section 3). Observed MGs are reported in
the literature by several authors \citep[e.g.,][] {Gor90, Dav92,
Car93, Meh03}. Classical MDC models \citep[e.g.,][]{Lar74} predict
$d(log\ Z)/d(log\ r)$ for ellipticals in the range $-0.5$ to
$-1.0$, while the MGs correlate with global properties of
elliptical galaxies, such that more massive galaxies have steeper
MGs. The observational results of correlations between MGs and
other internal global properties are controversial, as summarized
by \citet{Kob99}.

The scenario of HC suggests that present day galaxies form instead
by a sequence of mergers of smaller objects \citep[e.g.,][]
{Ste02}, and this is a natural consequence of the cold dark matter
theory of cosmological structure formation. These models have the
advantage of working in a cosmological context, explaining the
features of large scale structure. An important prediction of this
kind of model is that the MGs should be erased by the merger
process \citep[e.g.,][]{Whi80}. Flat MGs for some early-type
galaxies were reported by \citet{Car93}. Additional observational
evidence in favor of this scenario comes from the morphological
and kinematical disturbances found in some ellipticals such as
ripples/shells and multiple cores and kinematically decoupled
cores \citep[e.g.;][] {Sch92,deZ02}. The morphology-density
relation \citep{Dre80} can also be understood as the enhanced
formation of ellipticals in high density environments, places
where the efficiency of mergers is higher. In addition, the
observed growth of the number density of red galaxies from z
$\sim$ 1.4 to the present time \citep[]{Bel04,Fab05} argues in
favor of the HC scenario.

\citet{Kor89} considered a hybrid formation scenario for
early-types where both pictures should be combined to explain
observed structural properties of these types of galaxies.
Recently, \cite{Kob04} proposed the same idea based on simulations
involving formation and chemodynamical evolution of galaxies. She
shows that in a Cold Dark Matter scenario, the observed range of
MGs requires the occurrence of both mechanisms described above.
\citet{Kob04} suggests that by using the MG values, it is possible
to infer the merging histories of present-day galaxies.

The goal of this letter is to investigate the dependence of MGs on
$\sigma$ since this represents a powerful constraint on the
different scenarios outlined above and is tightly linked to the
mechanism of the formation of a galaxy. In Section 2, we present
the sample of early-types used and a brief description about the
method to measure MGs. In Section 3 we present and discuss our
results, summarizing our conclusions in Section 4.

%--------------------------------------------------------------

\section{The Sample and Line Index Determination}

Galaxies were selected from the ENEAR survey \citep{daC00}, which
contains a database of photometric \citep{Alo03} and spectroscopic
\citep{Weg03} parameters for a sample of early-type galaxies,
which are representative of the nearby Universe. Several global
parameters are available in the database such as magnitudes
($m_R$), effective radii ($r_{e}$), mean surface brightness within
$r_{e}$, characteristic diameter ($Dn$), velocity dispersion
($\sigma$) and $Mg_2$ line indices in the Lick system.

The spectroscopic data used in this letter were obtained with the
1.52 m telescope at the European Southern Observatory, during
several runs. The sample consists of 15 E and E/S0s and 14 S0
galaxies for which 1-D extractions were made as follows: the first
aperture (the central one) has 3 pixels ($\simeq$2.5"). Successive
lateral apertures are set in such a way that their central pixels
are the outermost pixels of the previous apertures, keeping the
same size. This process continues while a $S/N \geq20$ is obtained
for the spectra. When this condition fails, adjacent pixels
towards the external parts of the galaxy are added iteratively, up
to a maximum number of 5. The extraction process stops when it is
not possible to keep $S/N$ above 20, and this corresponds,
typically, to a region of the object of $\approx$ 0.5$r_e$. No
bias is expected in adopting variable radial zones as the
gradients are essentially linear over the considered radial range.
The final spectral resolution is $\sim4-5$ \AA. The $Mg_2$
measurements were converted into the Lick/IDS system and brought
to zero velocity dispersion following \citet{Wor97}.

A linear fit weighted by the index errors was used to measure its
variation as a function of $log(r/r_e^*)$, where $r_e^*$\ is
$r_e$\ corrected for the galaxy ellipticity
($r_e^*=r_e(1-\epsilon)^{-1/2}$). Because gradients are
systematically flattened within the inner $\sim2\arcsec$ due to
seeing effects, we excluded these inner regions from the fits. We
also discarded index measurements deviating by more than $2\sigma$
from the linear fits. A detailed description of our
data-reduction, analysis, and a comparison with gradient
determinations by other authors will be the subject of a
forthcoming paper. Meanwhile, a comparison of $Mg_2$ line
gradients obtained in this work for 5 galaxies (IC~2035, NGC~2663,
NGC~3557, NGC~3706 and NGC~5018) in common with \citet[]{Car93,
Car94b} shows good agreement. The mean difference between these
measurements is $-0.005\pm0.009$ mag dex$^{-1}$ which is quite
reassuring.

To perform the analysis described in the next section, we also
included the data obtained by \citet[]{Car93, Car94a, Car94b}. To
improve the homogeneity of the sample, we inspected visually the
morphological classification for all the objects. Some of the
\citet{Car93} $E$ and $E/S0$s were reclassified as $S0$s, based on
visual inspection of DSS B,R and I available images. We indicate
this whenever necessary in the text and figures. As a final remark
about the sample characteristics, we report that by means of a
Kolmogorov-Smirnov test, the set of galaxies used in this letter
has similar $\sigma$ distribution to that of ENEAR to a 99\%
confidence level.

%--------------------------------------------------------------

\section{Results and Discussion}

To test one of the most important predictions of the MDC model -
the dependence of MG with galaxy mass, we assume now that $Mg_2$
line index predominantly indicates metallicity. This line index
cannot, in general, be taken as indicator of the metallicity only,
since age effects should be considered. In fact, \citet{Meh03}
conclude that the presence of a MG and the absence of gradients in
age and $\alpha/Fe$ ratio for early-type galaxies in the Coma
cluster imply that the intrinsic $Mg-\sigma$ relation is driven by
metallicity alone. Nevertheless, we know that for our sample, made
up of galaxies from field and other clusters, some more broader
age distribution could play a role. While we are aware of these
caveats, we first discuss $Mg_2$ gradients assuming this index is
a good metallicity indicator. In order to further support our
results, we combine measurements of $Mg_2$ and $H\beta$ to infer
$[Z/H]$ gradients by comparison with the \citet{Tho03} models (see
below).

The essential result of this letter is shown in panel (a) of Fig.1
where we plot $\triangle Mg_2/ \triangle log\ r/r_e^*$ versus
$log\ \sigma$. Since there are few $E$s in the low $\sigma$ range,
we also included $S0$ galaxies, assuming that they have similar
metallicity properties as the $E$s \citep{Ber98}. This figure
shows that, while massive galaxies span a wide range of $Mg_2$
gradients there are no low mass galaxies with steep (very
negative) $Mg_2$ gradients. We point out that the lower envelope
of the data distribution seems to follow a linear relation between
$\triangle Mg_2/ \triangle log\ r/r_e^*$ and $\log \sigma$, which
indicates that the formation of the galaxies occupying that locus
might have been dominated by a monolithic collapse, as proposed by
\citet{Kob04}. In her model, galaxies predominantly formed by
monolithic collapse define a correlation with $\sigma$ extending
to the low mass domain as suggested by the plot. Since our sample
may be considered as representative of early-type galaxies in
general, the lack of objects with steep gradients in the low
$\sigma$ region is not a selection artifact, but instead a real
feature of the galaxy distribution in the $\triangle Mg_2/
\triangle log\ r/r_e^*$ versus $log\ \sigma$ diagram.

In panel (b) of Fig.1 we plot the $Mg_2$ gradient versus $log\
Mass$ ($Mass \propto r_e \sigma^2$) using $\sigma$ and $r_e$ from
our database. The distances were determined from the $D_n-\sigma$
relation \citep{Ber02}, adopting H$_0$ = 75 km s$^{-1}$. This plot
is very similar to that in panel (a), even though the $Mass$
estimate reflects the uncertainties of both, $\sigma$ and $r_e$,
coupled to the hypothesis that the objects obey the virial
theorem.

Using only $Mg_2$ to represent metallicity, we may neglect
possible age effects. Therefore, to make an estimate of the [Z/H]
gradients, we used $Mg_2$ and $H\beta$ measurements and single
stellar population models of \citet{Tho03} for [$\alpha$/Fe]=+0.3.
This value of $\alpha/Fe$ was found by \citet{Meh03} to be
representative of all spatial regions of their sample galaxies.
Metallicities were estimated for $r_e/8$ and $r_e/2$ from the
$Mg_2$ and $H\beta$ values interpolated in the linear index vs.
$\log r/r_e$ relations. Galaxies for which one or both points fell
out of the domain of the stellar population model predictions in
index-index plots were excluded from this exercise.

In Fig.2 we display the $\triangle [Z/H]/ \triangle log\ r/r_e^*$
versus $\sigma$ (panel a) and $\triangle [Z/H]/ \triangle log\
r/r_e^*$ versus $Mass$ (panel b) for 18 galaxies. We see that even
though the position of some galaxies relative to the others
changes substantially when we switch from $Mg_2$ to $[Z/H]$, the
overall distribution of the data points in Fig.2 is very similar
to that in Fig.1. In particular, low mass galaxies appear to be
characterized by low Z-gradients, whereas for high-mass galaxies
there is a large spread in this quantity. As suggested by the
simulations of \citet{Kob04} there seems to be a lower envelope in
the $\triangle [Z/H]$ versus $Mass$ plane which coincides with the
theoretical expectation from monolithic collapse models while the
upper envelope for massive galaxies coincides with products of
mergers.

The fact that there are three data points with small positive
gradients should not be taken at face value since they are
consistent with a flat profile if one takes into account the
associated errors. This should include errors such as H$\beta$
emission contamination, variations on the true $\alpha/Fe$ ratios
as well as the model uncertainties which are difficult to
estimate. Therefore, we consider our error bars as an
underestimate of the true error. Moreover, even if contrived,
there are models that could explain positive gradients, at least
for small mass objects \citep{Mor97}.

\citet{Gor90} and \citet{Dav92}, examining distinct samples of
early-types, found a loose correlation between MG and $\sigma$.
They claim that the strong scatter in that relation is a
consequence of different star formation histories. On the basis of
a larger sample, \citet{Car93} found that only low mass galaxies
follow a $MG-\sigma$ correlation, while \citet{Kob99} found no
such correlation.

Our results, on the other hand, show evidence of a dependence of
the $MG$ on $\sigma$ for the full range of $\sigma$ considered. In
other words, the plot of MG vs. $\sigma$ now includes new galaxies
for which a strong MG was found, providing a better definition of
the lower envelope of the distribution for large $\sigma$. It is
interesting to note, that although \citet{Kob04} simulations are
within the hierarchical formation scenario, her results show that
building galaxies from mergers of smaller objects with a high
fraction of gas mimics a monolithic collapse. One difference of
her simulations to a classic monolithic collapse is that the
timescales for the formation of her monolithic cases are a bit
longer.

A quantitative comparison between the MGs measured here and those
predicted by theoretical models \citep[e.g.,][]{Kob04,Pip04} is
made difficult by the fact that data and models sample different
spatial regions. For instance, \citet{Kob04} provide MG out to $2
r_e $ for objects in her simulations whereas our measurements
extend only to $\sim r_e/2$. Linear extrapolations of our
measurements to larger radial distances cannot be trusted, as real
galaxies could present substantial departures from linearity that
would lead to large errors. Nevertheless, a qualitative comparison
can be made between our results shown in Fig.2, with those of
simulations displayed in Fig.6 of \citet{Kob04}. The lower
boundary of the data points in our Fig.2 seems to harbor the
objects in Kobayashi's plot that were formed by monolithic
collapse and minor mergers, while the objects distributed in the
upper region (weak MGs) may represent those with important merger
histories.

\citet{Pip04} predict no correlation of MG with mass. In these
models the MG developed in a galaxy depends on the interplay
between star formation and stellar winds in the different galaxy
zones. The flat dependence of the MGs with mass in this case can
be explained by means of a strong ``inverse wind effect'' \citep
{Mat94}. These models reproduce well the increase of the
$\alpha/Fe$ ratio as a function of the galaxy mass by means of the
inverse-wind hypothesis. \citet{Mat94} showed that to explain the
increase of $\alpha/Fe$ in more massive galaxies it is important
to assume that they complete their formation before the less
massive ones. If this effect is strong enough, larger galaxies
stop their star formation very fast preventing the development of
strong gradients, leading to a flat dependence of MG with galaxy
Mass \citep{Pip04}. Another way in which models could produce both
$\alpha$-enhancement and stronger MGs in massive galaxies is via a
variable initial mass function, whereby more massive galaxies
would have IMFs biased towards high mass stars \citep{Mat94}.

%-------------------------------------------------------------

\section{Summary}

We analyzed MGs for a sample of early-type galaxies and our
results show: {\bf (a)} at least part of the objects follow the
predicted $MG-\sigma$ relation claimed to exist in the MDC
scenario, defining a lower boundary in a MG vs. $\sigma$ diagram;
{\bf (b)} the remaining objects may represent a sequence of
increasing relative importance of the HC process. Particularly,
higher mass galaxies scatter vertically in this diagram,
indicating a contribution by both formation processes. These
results give observational support to the hybrid scenario proposed
by \citet{Kor89} and \citet{Kob04} and show that we can interpret
galaxy formation not as an exclusive dominance of either MDC or HC
scenarios, but instead, both mechanisms contribute to the origin
of these objects. Each galaxy has its particular formation history
depending on the merger events of its building blocks, their
nature (predominantly gas or stars) and their efficiency to
collapse. In this case the location of a galaxy in a $MG-\sigma$
diagram may be a useful tool to infer the relative importance of
mergers or collapse to its formation.

Additional observations underway will be used to increase our
statistics and to cover more homogeneously the range of parameters
examined. This will allow us to investigate the existence of a
correlation between MGs and environmental density and to examine
the radial behavior of $\alpha/Fe$, to be presented in a future
paper.

%---------------------------------------------------------------
\acknowledgments

R.L.C.O. acknowledges CNPq Fellowship; M.A.G.M. CNPq grants
301366/86-1, 471022/03-9 and 305529/03-0; C.C. Financial support
from MIUR/COFIN 2003028039 and CNPq Fellowship; P.S.P. CNPq grant
301373/86-8 and CLAF; R.P.S. to financial support from HST
Treasury program grant GO-09455.05 to the University of Virginia.
The authors thank Charles Rit\'e and Christopher N.A. Willmer for
their contribution during an early phase of this project.

%----------------------------------------------------------------------

%-------------------------------------------------------------------------------
%\newpage
%-------------------------------------------------------------------------------

\begin{figure}
   \includegraphics[angle=-90,scale=0.6]{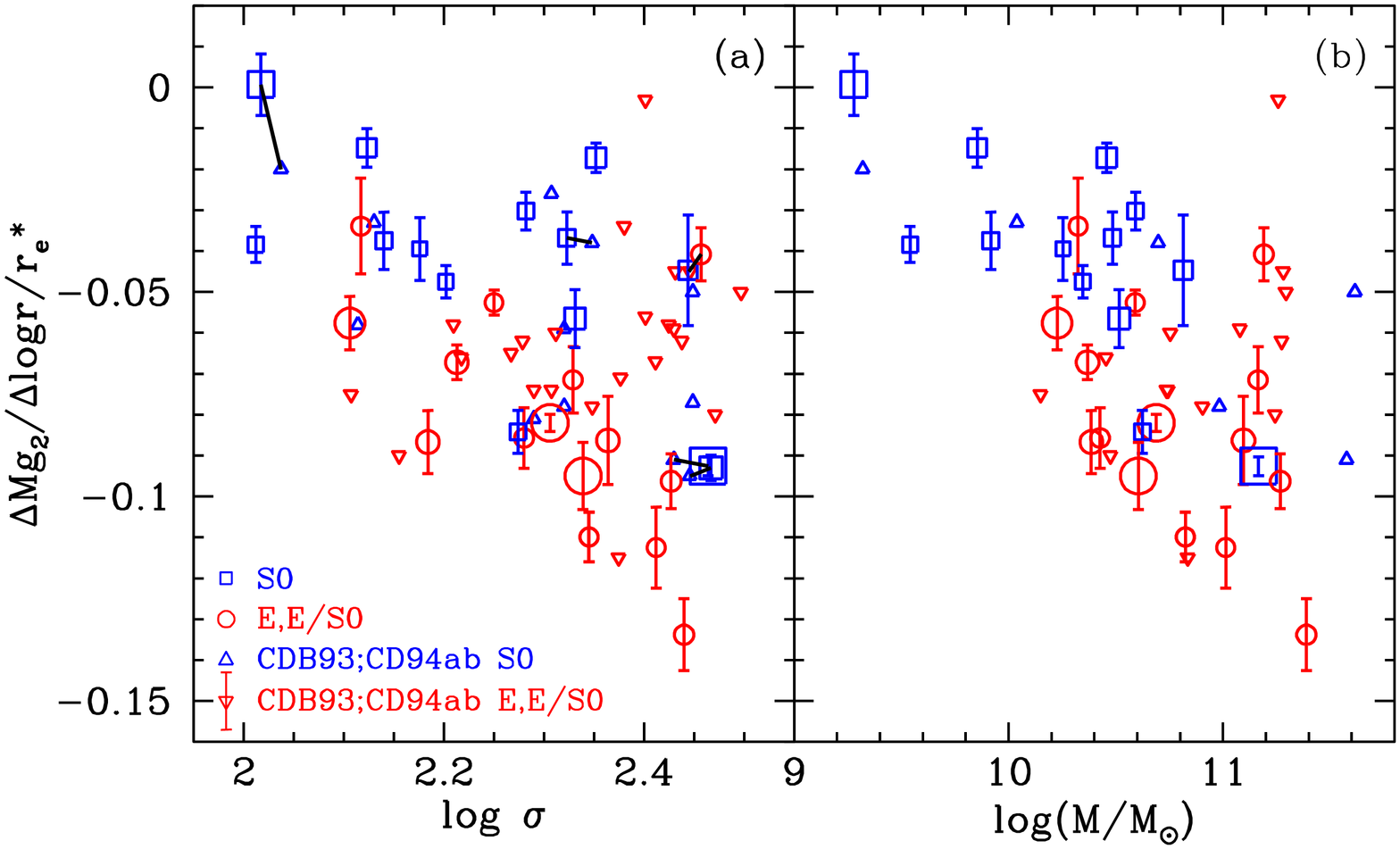}
\vspace{0.0 cm}
   \caption{Radial $Mg_2$ gradients as a function of central velocity dispersion
   $\sigma$ (panel a) and versus $Mass$ (panel b) for $E$, $E/S0$ and $S0$ galaxies.
   We also include the data for the early-types compiled by
   \citet[]{Car93, Car94a, Car94b} (CDB93+CD94ab).
   A few $E$s of these authors turned out to be $S0$s. Their $E$s and $E/S0$s are
   indicated by ``$\triangledown$'' and the $S0$s by ``$\vartriangle$''.
   The symbol ``$\triangledown$'' in the label inside the figure for
   CDB93+CD94a,b $E$, $E/S0$ contains an average error bar for their data.
   The 5 galaxies in common with their compilation are connected by a solid line
   to give an idea of the differences between our and their determinations of
   $\triangle Mg_2$ and $\sigma$.
   The size of the symbol is proportional to the fraction of $r_e$ in which we
   measured the $\triangle Mg_2$ (the bigger ones means $r/r_e\approx1$).
   }
   \label{fig1}
\end{figure}

%-------------------------------------------------------------------------------
\newpage
%-------------------------------------------------------------------------------

\begin{figure}
   \includegraphics[angle=-90,scale=0.6]{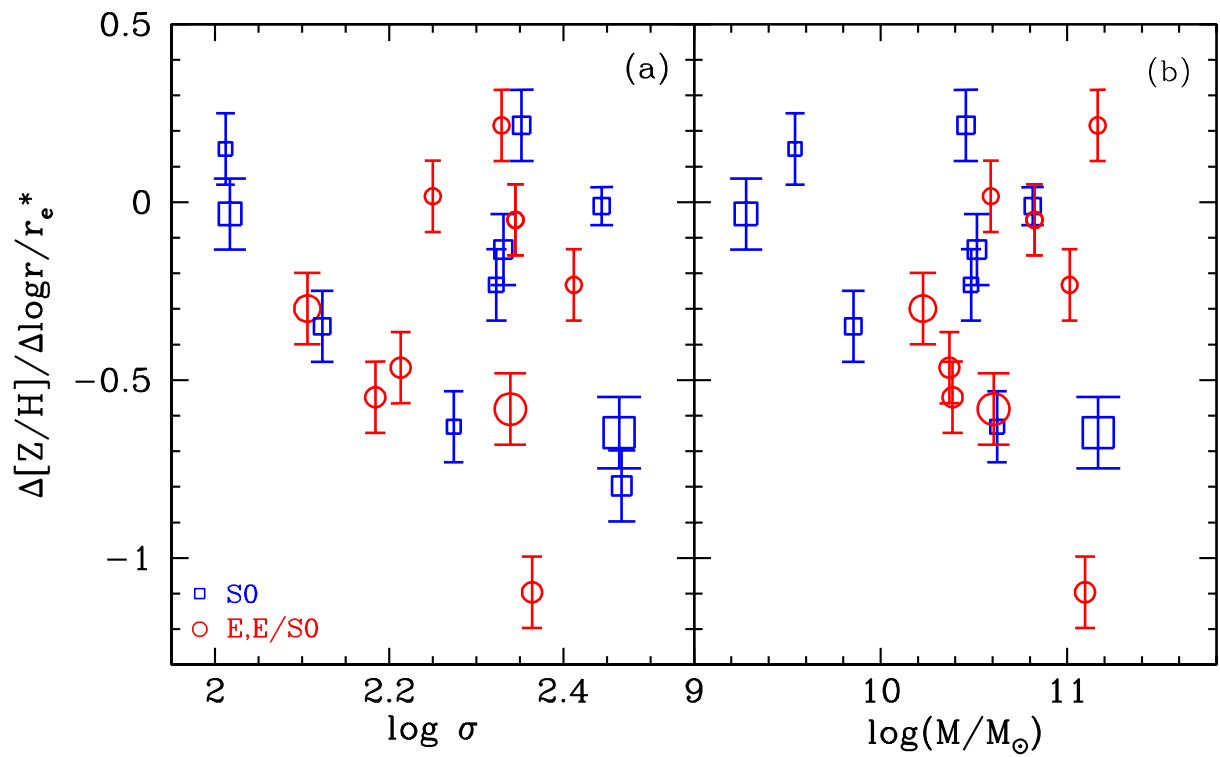}
\vspace{0.0 cm}
   \caption{Radial $\triangle [Z/H]$ as a function of galaxy central velocity
   dispersion $\sigma$ (panel a) and $Mass$ (panel b). The number of points is
   smaller in this plot because some line indices measurements of $Mg_2$ and
   $H\beta$ used to determine $[Z/H]$ fall outside the grid of single stellar
   population models by \citet{Tho03}.
   }
   \label{fig2}
\end{figure}

%-----------------------------------------------------------------------------

%------------------------------------------------------------------------------

\end{document}